\documentclass{JHEP3}

\usepackage{latexsym}
\usepackage{graphics}

\def\comment #1{}
\def\cf {{\it cf. }}

\def\refer #1{{(\ref{#1})}}
\def\fullref #1{\ref{#1} (p.\pageref{#1})}

\def\of #1{\!\left({#1}\right)}

\def\C {\mathrm{\,I\!\!\!C}}

\def\set #1{\left\lbrace{#1}\right\rbrace}

\def\brackets #1{\left[{#1}\right]}
\def\braces #1{\left\lbrace{#1}\right\rbrace}

\def\commutator #1#2{\brackets{{#1},{#2}}}
\def\antiCommutator #1#2{\braces{{#1},{#2}}}
\def\superCommutator #1#2{\brackets{{#1},{#2}}_\involution}

\def\defas {:=}

\def\equalby #1{\stackrel{\refer{#1}}{=}}

\def\involution {\mathbf \iota}

\def\endofproof {$\Box$\newline}

\def\eigenspace #1#2 {\mathrm{eig}\of{#1,#2}}

\newlength{\skiplength}
\def\skiph #1{\settowidth{\skiplength}{#1}\hspace{\skiplength}}

\def\inner {\!\cdot\!}

\title{DDF and Pohlmeyer invariants of (super)string}
\author{Urs Schreiber \\ Universit{\"a}t Duisburg-Essen \\ Essen, 45117, Germany\\
   E-mail: \email{Urs.Schreiber@uni-essen.de}}

\abstract{
We show how the
Pohlmeyer invariants of the bosonic string are expressible in terms of DDF invariants.
Quantization of the DDF observables in the usual way yields a
consistent quantization of the algebra of Pohlmeyer invariants. Furthermore it becomes
straightforward to generalize the Pohlmeyer invariants to the superstring as well as
to all backgrounds which allow a free field realization of the worldsheet theory.
}

\begin{document}

\newpage
\section{Introduction}

The classical string is known to be a completely integrable system with 
an infinite number of classical observables that Poisson-commute with all the
constraints. A concise and comprehensive review  of the work by Pohlmeyer,
Rehren, Bahns, et. al. 
\cite{Pohlmeyer:2002,MeusburgerRehren:2002,Pohlmeyer:1998,Pohlmeyer:1988}
on a particular manifestation of these \emph{gauge invariant observables}, known as 
\emph{Pohlmeyer invariants}, 
is given in \cite{Thiemann:2004}.

Since the Virasoro algebra is the direct sum of two copies
of $\mathrm{diff}(S^1)$, the diffeomorphism algebra of the circle, invariant
observables are simply those that are reparameterization invariant with respect
to these two algebras. Two common types of reparameterization invariant objects are
\begin{itemize}
  \item Wilson lines,
  \item integrals over densities of unit reparameterization weight .
\end{itemize} 
More precisely (see \S\fullref{Pohlmeyer invariants} 
for a detailed derivation), let $\mathcal{P}^\mu_\pm\of{\sigma}$ 
be the left- and right-moving classical fields 
on the free closed bosonic string with Poisson brackets of the form
\begin{eqnarray}
  \commutator{\mathcal{P}^\mu_\pm\of{\sigma}}{\mathcal{P}^\nu_\pm\of{\sigma^\prime}}_\mathrm{PB}
  &=&
  \pm \eta^{\mu\nu}\delta^\prime\of{\sigma,\sigma^\prime}
  \,.
\end{eqnarray}
These transform with unit weight under the action of the Virasoro algebra
and hence the \emph{Wilson line}
\begin{eqnarray}
  \mathrm{Tr}\mathbf{P}\exp\of{\int\limits_{S^1} \mathcal{P}^\mu_+  A_{+\mu}}
  \mathrm{Tr}\mathbf{P}\exp\of{\int\limits_{S^1} \mathcal{P}^\mu_-  A_{-\mu}}\,,
\end{eqnarray}
Poisson-commutes with all Virasoro constraints
(where $\mathbf{P}$ denotes path-ordering and 
$A_{\pm \mu}$ are two \emph{constant} Lie-algebra-valued 1-forms on target space). 
It is easy to see that also
the coefficients of $\mathrm{Tr}\of{A^n}$ in the Taylor expansion of this
object commute with the constraints. These coefficients are known as
the \emph{Pohlmeyer invariants}. The Poisson algebra of these
observables is rather convoluted. The problem of finding a quantum deformation of this
algebra turns out to be difficult and involved and has up to now remained
unsolved \cite{MeusburgerRehren:2002,Thiemann:2004,Bahns:2004}. 
Furthermore, by itself, it is not obvious how the above 
construction should generalize to the superstring.

For these reasons it seems worthwhile to consider the possibility
of alternatively using integrals over unit weight densities to construct a complete set
of classical invariant charges. A little reflection shows that the well-known
\emph{DDF operators} \cite{DelGiudiceDiVecchiaFubini:1972} for the 
covariantly quantized string,
which are operators that commute with all the quantum (super-)Virasoro constraints,
are built using essentially this principle:

From elementary CFT it follows that
for $\mathcal{O}(z)$ any primary CFT field of conformal weight $h=1$ we have
(after the usual introduction of a complex coordinate $z$ on the worldsheet)
\begin{eqnarray}
  \commutator{L_n}{\oint dz\, \mathcal{O}(z)} &=& 0\,, \hspace{.5cm}\forall\, n
  \,.
\end{eqnarray}
By choosing $\mathcal{O}(z) = \commutator{G_{-\nu}}{\tilde \mathcal{O}(z)}$,
(with $G_{-\nu}$ the $(\nu=0)$-mode (R sector) or $(\nu = -1/2)$-mode (NS sector) 
of the worldsheet supercurrent and 
$h\of{\tilde \mathcal{O}} = 1/2$) this generalizes to the superstring
\begin{eqnarray}
  \commutator{G_{n-\nu}}{\oint dz\, \mathcal{O}(z)} &=& 0 \,, \hspace{.5cm}\forall\, n
  \,,
\end{eqnarray}
as is reviewed below in \S\ref{DDF operators}. It hence only remains to find
$\mathcal{O}$ (or $\tilde \mathcal{O}$) of weight 1 (or 1/2) such that 
the resulting integrals have nice (super-)commutators and exhaust
the space of all invariant charges. Doing this in a natural way
yields the DDF operators. 

It is readily checked that this construction of the DDF operators can 
be mimicked in terms of the classical Poisson algebra to yield 
a complete set of classical invariants which we shall call
\emph{classical DDF invariants}. These inherit all the nice properties
of their quantum cousins. 

Most importantly, as is shown in \S\fullref{classical bosonic DDF invariants}, 
the Pohlmeyer invariants can be expressed in 
terms of the classical DDF invariants. Since it is known how the latter
have to be quantized (i.e. the crucial quantum corrections to these
charges is well known \cf \S\fullref{DDF operators}) 
this also tells us how the Pohlmeyer invariants
can consistently be quantized.

In particular this shows that any normal ordering in the quantization of the
Pohlmeyer invariants must be applied
only inside each DDF operator, while the DDF operators among themselves
need not be reordered. This clarifies the result of \cite{Bahns:2004},
where it was demonstrated that the Pohlmeyer invariants cannot be consistently
quantized by writing them in terms of worldsheet oscillators and
applying normal ordering with respect to these. Rather, as will be
shown here, one has to replace these oscillators with the corresponding
DDF observables, and the assertion is that the Pohlmeyer invariants, like
any other reparameterization invariant observable, are unaffected
by this replacement.

Because the DDF operators, together with the identity operator, form
a closed algebra, the quantization of the Pohlmeyer invariants in terms
of DDF operators, as demonstrated here, is manifestly consistent in the sense
that the quantum commutator of two such invariants is itself again an invariant.

It should be emphasized that, in contrast to what has been stated in 
\cite{Bahns:2004}, the construction of classical DDF invariants does \emph{not}
require that any worldsheet coordinate gauge has to be fixed, in particular
their construction has nothing to do with fixing conformal gauge. This
is obvious due to the fact that the classical DDF invariants are 
constructable (as the Pohlmeyer invariants, too) by proceeding from just the
Nambu-Goto action, which does not
even have an auxiliary worldsheet metric which could be gauge fixed. 
Furthermore the canonical data and the form of the Virasoro constraints as
obtained from the Nambu-Goto action are precisely the same as those
obtained from the Polyakov action with or without fixed worldsheet gauge.

Furthermore, our proof that the Pohlmeyer invariants can be equivalently
expressed in terms of DDF invariants (i.e. certain polynomials in DDF invariants are
equal to the Pohlmeyer invariants) constructively demonstrates that both are on the
same footing as far as requirements for their respective construction is concerned.\\

The organization of this paper is as follows:
  
We first review the construction of DDF operators in \S\fullref{DDF operators}
and then that of Pohlmeyer invariants in \S\fullref{Pohlmeyer invariants}.

Then in \S\fullref{classical bosonic DDF} we discuss the classical DDF invariants in detail, 
show how they can be used to 
express the Pohlmeyer invariants (\S\fullref{expressing Pohlmeyer invariants in terms of DDF invariants})
and how this generalizes to the superstring (\S\fullref{DDF and Pohlmeyer invariants for superstring}).

Summary and discussion is given in \S\fullref{summary and conclusion}.\\

A brief summary of the results presented here will be published in 
\cite{Schreiber:2004c}.

\section{DDF operators and Pohlmeyer invariants}
\label{DDF operators and Pohlmeyer invariants}

We review first the DDF operators, then the Pohlmeyer invariants, then show
how both are related.

\subsection{DDF operators }
\label{DDF operators}

The construction of DDF operators \cite{DelGiudiceDiVecchiaFubini:1972}
is very well known, but to the best of our knowledge there is no
comprehensive review of all possible cases (transversal and longitudinal,
bosonic and fermionic)
available in the standard literature. The following section
tries to list and derive all the essential facts.

In the standard textbook literature one can find 
\begin{itemize}
  \item in \cite{GreenSchwarzWitten:1987} (in non-CFT language)
the construction of 
\begin{itemize}
  \item transversal bosonic (\S 2.3.2) 
  \item transversal supersymmetric (\S 4.3.2) 
  \item longitudinal bosonic (pp. 111),
\end{itemize}
\item
and in \cite{Polchinski:1998} (in
CFT language) the construction of 
\begin{itemize}
  \item transversal bosonic (eq. (8.2.29)) 
\end{itemize}
DDF states, which go back to \cite{DelGiudiceDiVecchiaFubini:1972}.
\end{itemize}

The following summarizes and derives (in CFT language) \emph{all}
\begin{itemize}
  \item
    transversal and longitudinal
  \item
    bosonic and fermionic
\end{itemize}
DDF operators (for a free supersymmetric worldsheet theory).

Using the standard normalization of the OPE
\begin{eqnarray}
  X^\mu\of{z} X^\nu\of{0} 
  &\sim&
  -\frac{\alpha^\prime}{2}\eta^{\mu\nu}
  \ln z
  \nonumber\\
  \psi^\mu\of{z}\psi^\nu\of{0} &\sim&  \frac{\eta^{\mu\nu}}{z}
\end{eqnarray}
for the bosonic and fermionic worldsheet fields, the (super-)Virasoro currents read
\begin{eqnarray}
  T\of{z} &=& 
  -\frac{1}{\alpha^\prime} \partial X \inner \partial X\of{z}
  -\frac{1}{2}\psi \inner \partial \psi
  \nonumber\\
  T_{\rm F}\of{z}
  &=&
  i \sqrt{\frac{\alpha^\prime}{2}}
  \psi \inner \partial X
  \,.
\end{eqnarray}
The \emph{DDF operators} are defined as a set of operators that commute with all 
modes of $T$ and $T_\mathrm{F}$ (are 'gauge invariant observables') 
and satisfy an algebra that mimics that of worldsheet oscillator creation/annihilation operators.

First of all one needs to single out two linearly independent lightlike Killing vectors $l$ and $k$
on target space, 
and in the context of this subsection we choose to normalize them as $l\inner k = 2$. 
The span of $l$ and $k$ is called the \emph{longitudinal} space and its orthogonal
complement is the \emph{tranverse} space.

For $\mathcal{O}\of{z} = \sum\limits_{-\infty}^{\infty}\mathcal{O}_m z^{-(m+h)}$
a primary field of weight $h$  we shall refer to the OPE
$T\of{z}\mathcal{O}\of{0}\sim \frac{h}{z^2}\mathcal{O}\of{0} 
+ \frac{1}{z}\partial\mathcal{O}\of{0}$ as the \emph{tensor law} in some
of the following formulas, instead of writing out all terms.
The modes of $T$ and $T_\mathrm{F}$ are denoted by $L_m$ and $G_{m-\nu}$ as
usual.

The elementary but crucial fact used for the construction of DDF operators
is that 0-modes of tensor operators of
weight $h=1$ commute with all $L_m$ generators
according to
\begin{eqnarray}
  \label{tensor law}
  \commutator{L_m}{{\cal O}_n} &=& \left((h-1)m-n\right) {\cal O}_{m+n}
  \,.
\end{eqnarray}

Therefore the task of finding DDF states is reduced to that of finding 
linearly independent $h=1$  fields that have the desired commutation relations 
and, in the case of the superstring, are closed with respect to $T_\mathrm{F}$  
(see below).

\paragraph{Bosonic string.}

For the bosonic string the DDF operators $A^\mu_n$ are defined by
\begin{eqnarray}
  \label{bosonic string DDF operators}
  A^\mu_n
  &\propto&
  \oint \frac{dz}{2\pi i}
  \left(
    \partial X^\mu  
    + k^\mu\frac{\alpha^\prime}{8}in \partial \ln \left(k\inner \partial X\right) 
  \right)e^{in k\inner X}\of{z}
  \,.
\end{eqnarray}
(These are of course nothing but integrated vertex operators
of the massless fields.
Note that the logarithmic terms of $k \inner \partial X$, as well as the inverse powers
that will be used further below,
are well defined operators, as is discussed above equation (2.3.87) in \cite{GreenSchwarzWitten:1987}.)\\

It is straightforward to check that the operators \refer{bosonic string DDF operators} 
are really invariant:

First consider the transverse DDF operators.
For $v$ a transverse target space vector (such that in particular $v\inner k $ = 0 ) 
the operator $v\inner A_n$ is 
manifestly the 0-mode of an $h=1$ primary field 
(the exponential factor has $h=0$ due to $k\inner k = 0$) and hence is
invariant.

Furthermore 
$k \inner A_n \propto \delta_{n,0} k\inner \oint \partial X$ 
(for $n\neq 0$ the integrand is a total derivative)
also obvioulsy commutes with
the $L_m$. 

The only subtlety arises for the longitudinal $l\inner A_n$. Here,
the non-tensor behaviour of
\begin{eqnarray}
  T\of{z}  l\inner \partial X e^{in k\inner X}\of{w}
  &\sim&
  -\frac{\alpha^\prime}{2}\frac{in}{(z-w)^3}e^{in k\inner X}\of{w}
  +
  \mbox{$(h=1)$-tensor law}
\end{eqnarray}
is precisely canceled by the curious logarithmic correction term
$
  \partial \ln \left(k\inner \partial X\right) \of{z}
  =
  \frac{k\inner \partial^2 X}{k\inner \partial X}\of{z}
$.
Namely because of
\begin{eqnarray}
  T\of{z} \partial^2 X^\mu\of{w}
  &\sim&
  \frac{2 \partial X^\mu\of{w}}{(z-w)^3}
  +
  \mbox{$(h=2)$-tensor law}
\end{eqnarray}
one has
\begin{eqnarray}
  \Rightarrow
  T\of{z}
  \,
  \frac{k\inner \partial^2 X}{k\inner \partial X}e^{in k\inner X}\of{w}
  &\sim&
  \frac{2 e^{in k\inner X}}{(z-w)^3}
  +
  \mbox{$(h=1)$-tensor law}
  \,,
\end{eqnarray}
which hence makes the entire integrand of $l\inner A_m$ transform as an $h=1$  
primary, as desired.

\paragraph{Superstring.}
The analogous construction for the superstring has to ensure in addition that the DDF operators
commute with the supercharges $G_{m-\nu}$. 
This is simply achieved by `closing' the integral over 
a given weight $h=1/2$ primary field $D\of{z}$ to obtain the operator
\begin{eqnarray}
  \superCommutator{G_{-\nu}}{D_{\nu}}
  &=&
  \commutator{\oint \frac{dz}{2\pi i}T_F\of{z}}{\oint \frac{dz}{2\pi i}D\of{z}}
  \hspace{1cm}
  \left\lbrace
    \begin{array}{ll}
      \nu = 0 & \mbox{R sector}\\
      \nu = 1/2 & \mbox{NS sector}
    \end{array}
  \right.
  \,.
\end{eqnarray}
Here and in the remainder of this subsection the brackets denote supercommutators.

The resulting operator is manifestly the zero mode of a weight $h=1$ tensor and hence
commutes with all $L_n$. Furthermore it commutes with $G_{-\nu}$ because of
\begin{eqnarray}
  \superCommutator{G_{-\nu}}{\superCommutator{G_{-\nu}}{D_\nu}} 
  &=& 
  \superCommutator{L_{-2\nu}}{D_\nu}
  \;\stackrel{\refer{tensor law}}{=}\;
  0
  \,.
\end{eqnarray}
Since
$G_{-\nu}$ and $L_m, \forall\, m$ generate the entire algebra, the `closed' operator
$\commutator{G_{-\nu}}{D_\nu}$ indeed
commutes with all $L_m$ and $G_{m-\nu}\,,\forall\,m$.

It is therefore clear that the superstring DDF operators, which can be defined as
\begin{eqnarray}
  A_n^\mu
  &\defas&
  \commutator
  {G_\nu}
  {
     \oint \frac{dz}{2\pi i}
     \psi^\mu e^{i n k\inner X}\of{z} 
  }
  \nonumber\\
  B_n^\mu 
  &\defas&
  \commutator
  {G_\nu}
  {
    \oint \frac{dz}{2\pi i}
    \left(
      \psi^\mu \,k\inner \psi 
      -
      \frac{1}{4}
      k^\mu
      \partial \ln \left(k\inner \partial X\right)
    \right)
    \frac{e^{in k \inner X}}{\sqrt{k\inner \partial X}}
  }
\end{eqnarray}
commute with  the super-Virasoro generators, since the second arguments of the commutators
are integrals over weight 1/2 tensors. (And of course the latter are nothing but the 
integrated vertex operators
as they appear in the (-1) superghost picture). The nature and purpose of the logarithmic correction
term in the second line is just as discussed for the bosonic theory above: It cancels
the non-tensor term in
\begin{eqnarray}
  T\of{z}
  \,
  l\inner \psi\, k\inner\psi \frac{e^{i k\inner  X}}{\sqrt{k\inner \partial X}}\of{w}
  &\sim&
  \frac{1}{(z-w)^3}\frac{e^{i n k\inner X}}{\sqrt{k\inner \partial X}}
  +
  \mbox{$(h=1/2)$-tensor law}
  \,.
\end{eqnarray}

Evaluating the above supercommutators yields the explicit form for $A^\mu_n$ and $B^\mu_n$:
\begin{eqnarray}
  A^\mu_n &=&
  i \sqrt{\frac{2}{\alpha^\prime}}
  \oint \frac{dz}{2\pi i}
  \left(
    \partial X^\mu
    + 
    \frac{\alpha^\prime}{2}in \psi^\mu \, k\inner \psi
  \right)
  e^{ink\inner X}
  \of{z}
  \nonumber\\
  B^\mu_n
  &=&
  i \sqrt{\frac{2}{\alpha^\prime}}
  \oint \frac{dz}{2\pi i}
  \left(
    \partial X^\mu \, k\inner \psi
    -
    \psi^\mu k\inner \partial X
    +
    \frac{\alpha^\prime}{4}
    \psi^\mu \, k\inner \psi\, k\inner \partial \psi
    \frac{1}{k\inner \partial X}
  \right)
  \frac{e^{ink\inner X}}{\sqrt{k\inner \partial X}}
  \of{z}
  \nonumber\\
  &&+
  i \sqrt{\frac{2}{\alpha^\prime}}
  \oint \frac{dz}{2\pi i}
  k^\mu
  \left(
     k\inner \psi f_1\of{k\inner X,k\inner \partial X}
     +
     k\inner \partial \psi f_2\of{k\inner X, k\inner \partial X}
  \right)
  \of{z}
  \,,
\end{eqnarray}
where $f_1$ and $f_2$ are functions which we don not need to write out here.\\

The above discussion has focused on only a single chirality sector (left-moving, say).
It must be noted that the exponent $in k\inner X$ involved in the definition of
all the above DDF operators contains the 0-mode $k\inner x$ of the coordinate
field $k\inner X$. The existence of this 0-mode implies that the above DDF operators do
\emph{not} commute with the (super-)Virasoro generators of the opposite chirality.
In order to account for that one has to suitably multiply left- and right-moving DDF operators.
The details of this will be discussed in \S\fullref{classical bosonic DDF}.

\subsection{Pohlmeyer invariants}
\label{Pohlmeyer invariants}

We now turn to the classical bosonic string and discuss the
invartiants which have been studied by Pohlmeyer et al. 

In the literature the invariance of the Pohlmeyer charges is demonstrated
by the method of \emph{Lax pairs}. But the same fact follows already from the 
well-known reparameterization invariance property of Wilson loops.
To recall how this works for the classical bosonic string consider the following:

Denote the left- or rightmoving classical worldsheet fields in canonical language by
$\mathcal{P}^\mu\of{\sigma}$, which have the canonical Poisson bracket
\begin{eqnarray}
  \label{chiral CCR}
  \commutator{
    \mathcal{P}^\mu\of{\sigma}
  }
  {
    \mathcal{P}^\nu\of{\sigma^\prime}
  }_{\mathrm{PB}}
  &=&
  -
  \eta^{\mu\nu}
  \delta^\prime\of{\sigma-\sigma^\prime}
  \,.
\end{eqnarray}
The modes of the Virasoro constraints are
\begin{eqnarray}
  L_m &\defas&
  \frac{1}{2}
  \int d\sigma\,
  e^{-im\sigma}
  \eta_{\mu\nu}
  \mathcal{P}^\mu\of{\sigma}\mathcal{P}^\nu\of{\sigma}
\end{eqnarray}
and the $\mathcal{P}\of{\sigma}$ transform with unit weight under
their Poisson action:
\begin{eqnarray}
  \label{classical [Lm,Y]}
  \commutator{L_m}{\mathcal{P}^\mu\of{\sigma}}_{\mathrm{PB}}
  &=&
  \left(
    e^{-im\sigma}
    \mathcal{P}^\mu\of{\sigma}
  \right)^\prime
  \,.
\end{eqnarray}

This is all one needs to show that the \emph{Pohlmeyer invariants} 
$Z^{\mu_1 \cdots \mu_N}$ defined by
\begin{eqnarray}
  \label{definition Pohlmeyer invariants}
  Z^{\mu_1\cdots \mu_N}\of{\mathcal{P}}
  &:=&
  \frac{1}{N}
  \int\limits_0^{2\pi}
  d\sigma^1\,
  \int\limits_{\sigma^1}^{\sigma^1 + 2\pi}
  d\sigma^2\,  
  \cdots
  \int\limits_{\sigma^{N-1}}^{\sigma^1 + 2\pi}
  d\sigma^N\,  
    \mathcal{P}^{\mu_1}\of{\sigma^1}
    \mathcal{P}^{\mu_2}\of{\sigma^2}
    \cdots
    \mathcal{P}^{\mu_N}\of{\sigma^N}
  \nonumber\\
\end{eqnarray}
Poisson-commute with all the $L_m$.

The \emph{proof} involves just a little combinatorics and algebra:

First note that if 
$F\of{\sigma^1,\sigma^2,\cdots, \sigma^N}$
is any function which is periodic with period $2\pi$ in each
of its $N$ arguments, the cyclically permuted path-ordered
integral over $F$ is equal to the integral used in 
\refer{definition Pohlmeyer invariants}
\begin{eqnarray}
  &&
  \left[
    \int\limits_{0 < \sigma^1 < \sigma^2 < \cdots < \sigma^N < 2\pi}
    \!\!\!\!\!\!\!\!\!\!\!\!\!\!\!\!\!\!\!\!\!\!d^N \sigma
    \;\;\;\;\;\;\;+ 
    \int\limits_{0 < \sigma^N < \sigma^1 < \cdots < \sigma^{N-1} < 2\pi}
    \!\!\!\!\!\!\!\!\!\!\!\!\!\!\!\!\!\!\!\!\!\!d^N \sigma
    \;\;\;\;\;\;\;+
    \int\limits_{0 < \sigma^{N-1} < \sigma^N < \cdots < \sigma^{N-2} < 2\pi}
    \!\!\!\!\!\!\!\!\!\!\!\!\!\!\!\!\!\!\!\!\!\!d^N \sigma
  \;\;\;\;\;\;\;\right]  
  F\of{\sigma_1,\sigma_2,\cdots, \sigma_N}
  \nonumber\\
  &=&
  \int\limits_0^{2\pi}
  d\sigma^1\,
  \int\limits_{\sigma^1}^{\sigma^1 + 2\pi}
  d\sigma^2\,  
  \cdots
  \int\limits_{\sigma^{N-1}}^{\sigma^1 + 2\pi}
  d\sigma^N\,  
    F\of{\sigma_1,\sigma_2,\cdots ,\sigma_N}
  \,.
\end{eqnarray}
(This follows by noting that while, for instance, $\sigma^1$ runs from $0$ to $2\pi$
all other $\sigma^i$ can be taken to run from $\sigma^1$ to $\sigma^1 + 2\pi$ while
remaining in the correct order.)

This shows that the Pohlmeyer observables \refer{definition Pohlmeyer invariants} 
are 
invariant under cyclic permutation of their indices. It can also
be used to write their variation as
\begin{eqnarray}
  \delta Z^{\mu_1 \cdots \mu_N}
  &=&
  \frac{1}{N}
  \int\limits_0^{2\pi}
  d\sigma^1\,
  \int\limits_{\sigma^1}^{\sigma^1 + 2\pi}
  d\sigma^2\,  
  \cdots
  \int\limits_{\sigma^{N-1}}^{\sigma^1 + 2\pi}
  d\sigma^N\,  
  \Bigg(
    \mathcal{P}^{\mu_1}\of{\sigma^1}
    \mathcal{P}^{\mu_2}\of{\sigma^2}
    \cdots
    \delta\mathcal{P}^{\mu_N}\of{\sigma^N}  
    +
   \nonumber\\
   &&
   \skiph{$
  \frac{1}{N}
  \int\limits_0^{2\pi}
  d\sigma^1\,
  \int\limits_{\sigma^1}^{\sigma^1 + 2\pi}
  d\sigma^2\,  
  \cdots
  \int\limits_{\sigma^{N-1}}^{\sigma^1 + 2\pi}
  d\sigma^N\,  
  \Bigg(\;
$}
  +
    \mathcal{P}^{\mu_N}\of{\sigma^1}
    \mathcal{P}^{\mu_1}\of{\sigma^2}
    \cdots
    \delta\mathcal{P}^{\mu_{N-1}}\of{\sigma^N}+     
   \nonumber\\
   &&
   \skiph{$
  \frac{1}{N}
  \int\limits_0^{2\pi}
  d\sigma^1\,
  \int\limits_{\sigma^1}^{\sigma^1 + 2\pi}
  d\sigma^2\,  
  \cdots
  \int\limits_{\sigma^{N-1}}^{\sigma^1 + 2\pi}
  d\sigma^N\,  
  \Bigg(\;
$}
  + \cdots
   \Bigg)
  \,,
\end{eqnarray}
because we may cyclically permute the integration variables.
But if one now sets $\delta \mathcal{P}^\mu\of{\sigma} = 
\commutator{L_m}{\mathcal{P}^\mu\of{\sigma}}_\mathrm{PB}$
one gets, using \refer{classical [Lm,Y]},
\begin{eqnarray}
  &&
  \!\!\!\!\!\!\!\!\!\delta Z^{\mu_1 \cdots \mu_N} = 
  \nonumber\\
  &&
  \!\!\!\!\!\!\!
  \frac{1}{N}
  \int\limits_0^{2\pi}
  d\sigma^1\,
  \int\limits_{\sigma^1}^{\sigma^1 + 2\pi}
  d\sigma^2\,  
  \cdots
  \int\limits_{\sigma^{N-2}}^{\sigma^1 + 2\pi}
  d\sigma^{N-1}\,  
  \Bigg(
  \nonumber\\
  &&
    \xi\mathcal{P}^{\mu_N}
    \mathcal{P}^{\mu_1}\of{\sigma^1}
    \cdots
    \mathcal{P}^{\mu_{N-1}}\of{\sigma^{N-1}}
    -
    \mathcal{P}^{\mu_1}\of{\sigma^1}
    \cdots
    \xi
    \mathcal{P}^{\mu_{N-1}}
    \mathcal{P}^{\mu_N}\of{\sigma^{N-1}}
    +
   \nonumber\\
   &&
  +
    \xi
    \mathcal{P}^{\mu_{N-1}}
    \mathcal{P}^{\mu_N}\of{\sigma^1}
    \cdots
    \mathcal{P}^{\mu_{N-2}}\of{\sigma^{N-1}}
    -
    \mathcal{P}^{\mu_N}\of{\sigma^1}
    \cdots
    \xi
    \mathcal{P}^{\mu_{N-2}}
    \mathcal{P}^{\mu_{N-1}}\of{\sigma^{N-1}}
   \nonumber\\
   &&
   \!\!\!\!\!\!\! 
   +
   \cdots
   \Bigg)
  \nonumber\\
  && 
  \!\!\!\!\!\!\!\!\! = 0
\end{eqnarray}
(where we have written $\xi\of{\sigma} = e^{-im\sigma}$ for brevity).
The contributions from the innermost integration cancel due to the cyclic
permutation of integrands and integration variables.
\endofproof

We note that the identity 
$\commutator{L_m}{Z^{\mu_1 \cdots \mu_N}\of{\mathcal{P}}}_\mathrm{PB} = 0$
is just the infinitesimal version of the fact that the Pohlmeyer observables
are invariant under \emph{finite reparameterizations}
\begin{eqnarray}
  \label{reparameterized cal P}
  \mathcal{P}\of{\sigma}
  &\mapsto&
  \tilde \mathcal{P}\of{\sigma}
  \defas
  R^\prime\of{\sigma}
  \mathcal{P}\of{R\of{\sigma}}
\end{eqnarray}
induced by the \emph{invertible} function $R$ which is assumed to satisfy
\begin{eqnarray}
  \label{cyclic property of R}
  R\of{\sigma+2\pi} = R\of{\sigma} + 2\pi
  \,.
\end{eqnarray}

Indeed, we have the important relation
\begin{eqnarray}
  \label{finite invariance}
  Z^{\mu_1 \cdots \mu_N}\of{\mathcal{P}}
  &=&
  Z^{\mu_1 \cdots \mu_N}\of{\tilde \mathcal{P}}
  \,,
\end{eqnarray}
which is at the heart of our derivation in \S\fullref{expressing Pohlmeyer invariants in terms of DDF invariants} 
that the Pohlmeyer invariants can be expressed in terms of DDF invariants

The \emph{proof} of this involves just a simple change of variables in the 
integral:
\begin{eqnarray}
  \label{proof of finite invariance}
  &&Z^{\mu_1 \cdots \mu_N}\of{\tilde \mathcal{P}}
  \nonumber\\
  &=&
  \frac{1}{N}
  \int\limits_0^{2\pi}
  d\sigma^1\,
  \int\limits_{\sigma^1}^{\sigma^1 + 2\pi}
  d\sigma^2\,  
  \cdots
  \int\limits_{\sigma^{N-1}}^{\sigma^1 + 2\pi}
  d\sigma^N\,  
    R^\prime\of{\sigma^1}R^\prime\of{\sigma^2}\cdots R^\prime\of{\sigma^N}
    \mathcal{P}^{\mu_1}\of{R\of{\sigma^1}}
    \cdots
    \mathcal{P}^{\mu_N}\of{R\of{\sigma^N}}  
  \nonumber\\
  &\stackrel{\tilde \sigma^i \defas R\of{\sigma^i}}{=}&
  \frac{1}{N}
  \int\limits_{R\of{0}}^{R\of{2\pi}}
  d\tilde \sigma^1\,
  \int\limits_{R\of{\sigma^1}}^{R\of{\sigma^1 + 2\pi}}
  d\tilde \sigma^2\,  
  \cdots
  \int\limits_{R\of{\sigma^{N-1}}}^{R\of{\sigma^1 + 2\pi}}
  d\tilde \sigma^N\,  
    \mathcal{P}^{\mu_1}\of{\tilde \sigma^1}
    \mathcal{P}^{\mu_2}\of{\tilde \sigma^2}
    \cdots
    \mathcal{P}^{\mu_N}\of{\tilde \sigma^N}  
  \nonumber\\
  &\equalby{cyclic property of R}&
  \frac{1}{N}
  \int\limits_{R\of{0}}^{R\of{0}+2\pi}
  d\tilde \sigma^1\,
  \int\limits_{\tilde \sigma^1}^{\tilde \sigma^1 + 2\pi}
  d\tilde \sigma^2\,  
  \cdots
  \int\limits_{\tilde\sigma^{N-1}}^{\tilde \sigma^1 + 2\pi}
  d\tilde \sigma^N\,  
    \mathcal{P}^{\mu_1}\of{\tilde \sigma^1}
    \mathcal{P}^{\mu_2}\of{\tilde \sigma^2}
    \cdots
    \mathcal{P}^{\mu_N}\of{\tilde \sigma^N}
  \nonumber\\
  &=&  
  Z^{\mu_1 \cdots \mu_N}\of{\mathcal{P}}
  \,.
\end{eqnarray}
\endofproof

Finally, for the sake of completeness, 
we note the well-known fact that the Pohlmeyer invariants appear naturally as the 
Taylor-coefficients of
\emph{Wilson loops} along the string at constant worldsheet time.
Let $A_\mu$ be a  \emph{constant but otherwise arbitrary} $\mathrm{GL}\of{N,\C}$ 
connection on target space, then the Wilson loop around the string of this connection with respect
to $\mathcal{P}$ is
\begin{eqnarray}
  {\rm Tr}\,
  \mathbf{P}
  \exp\of{
    \int\limits_0^{2\pi}
    d\sigma\,
    A_\mu \mathcal{P}^\mu\of{\sigma}
  }
  &=&
  \sum\limits_{n=0}^\infty
  Z^{\mu_1 \cdots \mu_n}\of{\mathcal{P}}
  \,
  \mathrm{Tr}
  \of{A_{\mu_1}\cdots A_{\mu_2}}
  \,,
\end{eqnarray}
where $\mathbf{P}$ denotes path-ordering.

This way of getting string ``states'' by means of Wilson lines
of constant (but possibly large $N$) gauge connections 
along the string is intriguingly reminiscent of similar constructions 
used in the
IIB Matrix Model (IKKT model) \cite{AokiIsoKawaiYoshihisaKitazawaTsuchiyaTada:1999}.\\

In the next sections 
the classical DDF invariants are described and it is shown how the Pohlmeyer
invariants can be expressed in terms of these.

\subsection{Classical bosonic DDF invariants and their relation to the Pohlmeyer invariants}
\label{classical bosonic DDF invariants}

The construction of classical DDF-like invariants for the
\emph{super}string, which is the content of
\S\fullref{DDF and Pohlmeyer invariants for superstring},
is straightforward once the bosonic case is understood. 
The basic idea is very simple and shall 
therefore be given here first for
the bosonic string, in order to demonstrate how 
\S\fullref{DDF operators} and \S\fullref{Pohlmeyer invariants} 
fit together.

\subsubsection{Classical bosonic DDF invariants}
\label{classical bosonic DDF}

In order to establish our notation and sign conventions we 
briefly list some definitions and relations which are in principle 
well known from elementary CFT but are rarely written out in the 
canonical language which we will need here.

So let $X\of{\sigma}$ and $P\of{\sigma}$ be canonical coordinates and
momenta of the bosonic string with Poisson brackets
\begin{eqnarray}
  \commutator{X^\mu\of{\sigma}}{P_\nu\of{\kappa}}_\mathrm{PB}
  &=&
  \delta^\mu_\nu\, \delta\of{\sigma-\kappa}
  \,.
\end{eqnarray}

In close analogy to the CFT notation $\partial X$ and $\bar \partial X$ 
we define
\begin{eqnarray}
  \mathcal{P}_\pm^\mu(\sigma)
  &=&
  \frac{1}{\sqrt{2T}}\left(
    P^\mu\of{\sigma} \pm T X^{\prime \mu}\of{\sigma}
  \right)
  \,.
\end{eqnarray}
(Here $T = 1/2\pi \alpha^\prime$ is the string tension and we assume a trivial
Minkowski background and shift all spacetime indices with 
$\eta_{\mu\nu} = \mathrm{diag}(-1,1,\cdots,1)$.)

Their Poisson brackets are of course
\begin{eqnarray}
  \commutator{\mathcal{P}_\pm^\mu\of{\sigma}}{\mathcal{P}_\pm^\nu\of{\kappa}}_\mathrm{PB}
  &=&
  \pm \eta^{\mu\nu}\delta^\prime\of{\sigma-\kappa}
  \nonumber\\
  \commutator{\mathcal{P}_\pm^\mu\of{\sigma}}{\mathcal{P}_\mp^\nu\of{\kappa}}_\mathrm{PB}
  &=&
  0
  \,.
\end{eqnarray}
From the mode expansion
\begin{eqnarray}
  \label{oscillator expansion}
  \mathcal{P}^\mu_+\of{\sigma}
  &\defas&
  \frac{1}{\sqrt{2\pi}}\sum_m \tilde \alpha_m^\mu e^{-im\sigma}
  \nonumber\\
  \mathcal{P}^\mu_-\of{\sigma}
  &\defas&
  \frac{1}{\sqrt{2\pi}}\sum_m \alpha_m^\mu e^{+im\sigma}
\end{eqnarray}
one finds the string oscillator Poisson algebra
\begin{eqnarray}
  \commutator{\alpha_m^\mu}{\alpha_n^\nu}_\mathrm{PB}
  &=&
  -i\, m\, \eta^{\mu\nu}\delta_{m+n,0}
  \,,
\end{eqnarray}
as well as
\begin{eqnarray}
  \commutator{x^\mu}{p^\nu}_\mathrm{PB} 
  &=&
  \eta^{\mu\nu}
  \,,
\end{eqnarray}
where
\begin{eqnarray}
  x^\mu &\defas& \frac{1}{2\pi}\int X^\mu\of{\sigma}\,d\sigma
  \nonumber\\
  p^\mu &\defas& \int P^\mu\of{\sigma}\, d\sigma \;=\; \frac{1}{\sqrt{4\pi T}}\alpha_0 
    = \frac{1}{\sqrt{4\pi T}}\tilde \alpha_0
  \,.
\end{eqnarray}

In terms of these oscillators the field $X^\prime$ reads
\begin{eqnarray}
  X^{\prime\mu}\of{\sigma}
  &=&
  \frac{1}{\sqrt{2T}}
  \left(
    \mathcal{P}_+^\mu\of{\sigma}
    -
    \mathcal{P}_-^\mu\of{\sigma}
  \right)
  \nonumber\\
  &=&
  \frac{1}{\sqrt{4\pi T}}
  \sum\limits_{m=-\infty}^\infty
  \left(
    -\alpha^\mu_m + \tilde \alpha_{-m}^\mu
  \right)
  e^{+im\sigma}
\end{eqnarray}
and hence the canonical coordinate field itself is
\begin{eqnarray}
  X^\mu\of{\sigma}
  &=&
  x^\mu +
  \frac{i}{\sqrt{4\pi T}}
  \sum\limits_{m\neq 0}
  \frac{1}{m}
  \left(
    \alpha_m^\mu - \tilde \alpha_{-m}^\mu
  \right)
  e^{+im\sigma}
  \,.
\end{eqnarray}

Any field $A\of{\sigma}$ is said to have \emph{classical conformal weight} $w\of{A}$ iff
\begin{eqnarray}
  \commutator{L_m}{A\of{\sigma}}
  &=&
  e^{-im\sigma}A^\prime\of{\sigma}
  +
  w\of{A}(e^{-im\sigma})^\prime A\of{\sigma}
\end{eqnarray}
and is said to have classical conformal weight $\tilde w\of{A}$ iff
\begin{eqnarray}
  \commutator{\tilde L_m}{A\of{\sigma}}
  &=&
  -e^{+im\sigma}A^\prime\of{\sigma}
  -
  \tilde w\of{A}(e^{+im\sigma})^\prime A\of{\sigma}
  \,,
\end{eqnarray}
where 
\begin{eqnarray}
  L_m &\defas&
  \frac{1}{2}\int e^{-im\sigma} \mathcal{P}_-\of{\sigma} \inner \mathcal{P}_-\of{\sigma}
  \;=\;
  \frac{1}{2}
  \sum\limits_{k=-\infty}^\infty
  \alpha_{m-k}\inner \alpha_k
  \nonumber\\
  \tilde L_m &\defas&
  \frac{1}{2}\int e^{+im\sigma} \mathcal{P}_+\of{\sigma} \inner \mathcal{P}_+\of{\sigma}  
  \;=\;
  \frac{1}{2}
  \sum\limits_{k=-\infty}^\infty
  \tilde \alpha_{m-k}\inner \tilde \alpha_k  
\end{eqnarray}
are the usual modes of the Virasoro generators.

The parts of $X\of{\sigma}$ which have $w = 0$ and $\tilde w = 0$, respectively, are
\begin{eqnarray}
  X^\mu_-\of{\sigma}
  &\defas&
  x^\mu
  -
  \frac{\sigma}{4\pi T}p^\mu
  +
  \frac{i}{\sqrt{4\pi T}}\sum_{m\neq 0} \frac{1}{m}\alpha_m^\mu e^{+im\sigma}
\end{eqnarray}
and
\begin{eqnarray}
  X^\mu_+\of{\sigma}
  &\defas&
  x^\mu
  +
  \frac{\sigma}{4\pi T}p^\mu
  +
  \frac{i}{\sqrt{4\pi T}}\sum_{m\neq 0} \frac{1}{m}\tilde \alpha_m^\mu e^{-im\sigma}
  \,.
\end{eqnarray}
This is checked by noticing the crucial property
\begin{eqnarray}
  \left(X^{\mu}_-\right)^\prime\of{\sigma}
  &=&
  -\frac{1}{\sqrt{2T}}
  \mathcal{P}^\mu_-\of{\sigma}  
  \nonumber\\
  \left(X^{\mu}_+\right)^\prime\of{\sigma}
  &=&
  \frac{1}{\sqrt{2T}}
  \mathcal{P}^\mu_+\of{\sigma}
  \,.  
\end{eqnarray}

These weight 0 fields can now be used to construct ``invariant oscillators'', namely the 
classical DDF invariants:

To that end fix any lightlike vector field $k$ on target space and consider the fields
\begin{eqnarray}
  R_\pm\of{\sigma}
  &\defas&
  \pm
  \frac{4\pi T}{k\inner p}\,
    k\inner X_\pm\of{\sigma}
  \,.
\end{eqnarray}
The prefactor is an invariant and chosen so that
\begin{eqnarray}
  R_\pm\of{\sigma + 2\pi}
  &=&
  R_\pm\of{\sigma} + 2\pi
  \,.
\end{eqnarray}
Furthermore the derivative of $R_\pm$ is
\begin{eqnarray}
  R^\prime_\pm\of{\sigma}
  &=&
  \frac{2\pi\sqrt{2T}}{k\inner p}\, k\inner \mathcal{P}_\pm\of{\sigma}
  \,.
\end{eqnarray}
It has been observed \cite{Rehren:2004} that this derivative vanishes only on a
subset of phase space of vanishing measure. This can be seen as follows:

The classical Virasoro constraints $\mathcal{P}_\pm^2 = 0$ say that $\mathcal{P}_\pm\of{\sigma}$ 
is lightlike. Because $k$ is also lightlike this implies that $k\inner \mathcal{P}_\pm\of{\sigma}$
vanishes iff $\mathcal{P}_\pm\of{\sigma}$ is parallel to $k$. 

By writing $\mathcal{P}_\pm = \mathcal{P}_\pm^0\left[1,\mathcal{P}_\pm^i/\mathcal{P}_\pm^0\right]$
and noting that the spatial unit vector $e^i_\pm\of{\sigma} \defas
\mathcal{P}_\pm^i/\mathcal{P}_\pm^0$ is of weight $w = 0$ or $\tilde w = 0$
(while it Poisson commutes with the respective opposite Virasoro algebra), and
hence transforms under the action of the Virasoro generators (which includes time evolution)
as $e^i_\pm\of{\sigma} \to e^i_\pm\of{\sigma + f\of{\sigma}}$, one sees that this condition
is satisfied for some $\sigma$ at some instance of time if and only if it is satisfied for some $\sigma$
at any given time. In other words the time evolution of the string traces out trajectories
in phase space which either have $\mathcal{P}_\pm$ parallel to $k$ for some $\sigma$ 
at \emph{all times} or \emph{never}.

In summary this means that except on the subset of phase space (of vanishing measure) 
of those trajectories where there exists a $\sigma$ such that
$k\inner \mathcal{P}_\pm\of{\sigma} = 0$, the observables $R_\pm\of{\sigma}$ define invertible
reparameterizations of the interval $[0,2\pi)$, as considered in \refer{cyclic property of R}.

The above fact will be crucial below for expressing the Pohlmeyer invariants in terms of DDF invariants.
For later usage 
let us introduce the notation $\mathbf{P}_k$ for the total phase space minus that set of 
vanishing measure:
\begin{eqnarray}
  \label{nice part of phase space}
  \mathbf{P}_k
  &\defas&
  \set{
    (X\of{\sigma},P\of{\sigma})_{\sigma \in (0,2\pi)}
    |
    k\inner\mathcal{P}_\pm\of{\sigma} \neq 0\;\forall \sigma
  }
  \,.
\end{eqnarray}

Now the classical DDF observables $A_m^\mu$ and $\tilde A_m^\mu$ of the closed bosonic string are finally defined 
(adapting the construction of \refer{bosonic string DDF operators} but using slighly different
normalizations)
by
\begin{eqnarray}
  \label{definition classical DDF}
  A_m^\mu
  &\defas&
  \frac{1}{\sqrt{2\pi}}
  \int d \sigma\,
  \mathcal{P}_-^\mu\of{\sigma}
  e^{
    -im R_-\of{\sigma}
  }
  \nonumber\\
  \tilde A_m^\mu
  &\defas&
  \frac{1}{\sqrt{2\pi}}
  \int d \sigma\,
  \mathcal{P}_+^\mu\of{\sigma}
  e^{
    im R_+\of{\sigma}
  }
  \,.
\end{eqnarray} 
Note that the construction principle of these objects is 
essentially the same as that of the ordinary oscillators \refer{oscillator expansion}
except that the parameterization of the string used here differs from one point in phase space to 
the other.

Being integrals over fields of total weight $w=1$ and $\tilde w = 1$, respectively, the 
DDF observables obviously Poisson-commute with their associated \emph{half} of the Virasoro generators:
\begin{eqnarray}
  \label{invariance pure}
  \commutator{L_m}{A_n^\mu} &=& 0
  \nonumber\\
  \commutator{\tilde L_m}{\tilde A_n^\mu} &=& 0
  \,.
\end{eqnarray}

But due to the coordinate 0-mode 
$\frac{2T}{k\inner p}\,k\inner x$
that enters the definition of
$R_\pm$, the mixed Poisson-brackets do not vanish. In order to construct invariants one therefore
has to split off this 0-mode and define the truncated observables
\begin{eqnarray}
  a^\mu_m &\defas& A^\mu_m e^{-im \frac{2T}{k\inner p}\,k\inner x}
  \nonumber\\
  \tilde a^\mu_m &\defas& A^\mu_m e^{-im \frac{2T}{k\inner p}\,k\inner x}
  \,.
\end{eqnarray}
These now obviously have vanishing \emph{mixed} Poisson brackets:
\begin{eqnarray}
  \label{invariance mixed}
  \commutator{L_m}{\tilde a_n^\mu} &=& 0
  \nonumber\\
  \commutator{\tilde L_m}{a_n^\mu} &=& 0
  \,.  
\end{eqnarray}

Therefore classical DDF invariants which Poisson commute with \emph{all}
Virasoro constraints are obtained by forming products 
\begin{eqnarray}
  \label{DDF invariants}
  D_{\set{m_i,\tilde m_j}} 
  &\defas& 
  a^{\mu_1}_{m_1}\cdots a^{\mu_r}_{m_r}\,\tilde a^{\nu_1}_{\tilde n_1}\cdots a^{\nu_s}_{\tilde m_s}
  e^{i N \frac{2T}{k\inner p}\, k\inner x}
\end{eqnarray}
which satisfy the \emph{level matching condition}:
\begin{eqnarray}
  \label{level matching condition}
  \sum\limits_i m_i \;=\; N \;=\;  \sum\limits_j \tilde m_j
  \,.
\end{eqnarray}

In order to see this explicitly write
\begin{eqnarray}
  \commutator{L_n}{
  D_{\set{m_i,\tilde m_j}}
  }_\mathrm{PB}
  &=& 
  \underbrace{
  \commutator{L_n}
  {
  a^{\mu_1}_{m_1}\cdots a^{\mu_r}_{m_r}e^{i N \frac{2T}{k\inner p}\,k\inner x}
  }_\mathrm{PB}
  }_{\equalby{invariance pure} 0}
  \tilde a^{\nu_1}_{\tilde n_1}\cdots a^{\nu_s}_{\tilde m_s}
  +
  \nonumber\\
  &&+
  a^{\mu_1}_{m_1}\cdots a^{\mu_r}_{m_r}e^{i N \frac{2T}{k\inner p}\,k\inner x}
  \underbrace{
  \commutator{L_n}{
    \tilde a^{\nu_1}_{\tilde n_1}\cdots a^{\nu_s}_{\tilde m_s}
  }_\mathrm{PB}
  }_{\equalby{invariance mixed} 0}
  \nonumber\\
  \commutator{\tilde L_n}{
  D_{\set{m_i,\tilde m_j}}
  }_\mathrm{PB}
  &=& 
  \underbrace{
  \commutator{\tilde L_n}
  {
  a^{\mu_1}_{m_1}\cdots a^{\mu_r}_{m_r}
  }_\mathrm{PB}
  }_{\equalby{invariance mixed} 0}
  \tilde a^{\nu_1}_{\tilde n_1}\cdots a^{\nu_s}_{\tilde m_s}e^{i N \frac{2T}{k\inner p}\,k\inner x}
  +
  \nonumber\\
  &&+  a^{\mu_1}_{m_1}\cdots a^{\mu_r}_{m_r}
  \underbrace{
  \commutator{L_n}{
    \tilde a^{\nu_1}_{\tilde n_1}\cdots a^{\nu_s}_{\tilde m_s}e^{i N \frac{2T}{k\inner p}\,k\inner x}
  }_\mathrm{PB}
  }_{\equalby{invariance pure} 0}
  \,.
\end{eqnarray}

This establishes the classical invariance of the DDF observables $D_{\set{m_i,\tilde m_j}}$.
We next demonstrate how the Pohlmeyer invariants can be expressed in terms of DDF invariants.

\subsubsection{Expressing Pohlmeyer invartiants in terms of DDF invariants}
\label{expressing Pohlmeyer invariants in terms of DDF invariants}

From the Fourier mode-like objects $A_m^\mu$ and $\tilde A_m^\mu$ one reobtains
quasi-local fields\footnote{
  It is interesting to discuss these fields, and in particular their
  quantization, from the point of view of
  worldsheet (quantum) gravity:

    Clearly the $\mathcal{P}_\pm^\mu\of{\sigma}$ are `not physical'
  (do not Poisson commute with the constraints) because they evaluate the
string's momentum and tension energy at a given value of the parameter $\sigma$, which 
of course has no physical relevance whatsoever. Heuristically, a physical observable
may make recourse only to values of fields of the theory, not to values of
auxiliary unphysical parameters. That is precisely the role played by the fields $R_\pm$.
They allow to characterize a point of the string purely in terms of physical
fields (string oscillations). Instead of asking: ``What is the value of $\mathcal{P}_\pm$
at $\sigma = 3$?'', we may ask the physically meaningful question:
``What is the value of $\mathcal{P}_\pm$ at a
point on the string where its configuration is such that $R_+ = 3$?''	
Quasi-local observables like the $\mathcal{P}^R_\pm$, or rather their absence,
are related to old and well known issues of (quantum) gravity in higher dimensions, 
often referred to in the context of ``\emph{the problem of time}'' \protect\cite{Carlip:2001}.

It is maybe instructive to note how these issues are resolved here for the 
\emph{worldsheet} theory of the relativistic string,
a toy example for quantum gravity when regarded as a theory of 1+1 dimensional gravity.
(Of course the string is rather more than a toy example for quantum gravity from the
\emph{target space} perspective.)
} $\mathcal{P}^R_\pm$ by an inverse Fourier transformation:
\begin{eqnarray}
  \label{local pre-invariants}
  \mathcal{P}^R_-\of{\sigma}
  &\defas&
  \frac{1}{\sqrt{2\pi}}
  \sum\limits_m A_m^\mu e^{+im\sigma}
  \;=\;
  \left((R_-)^{-1}\right)^\prime\of{\sigma}
  \mathcal{P}^\mu\of{(R_-)^{-1}\of{\sigma}}
  \nonumber\\
  \mathcal{P}^R_+\of{\sigma}
  &\defas&
  \frac{1}{\sqrt{2\pi}}
  \sum\limits_m \tilde A_m^\mu e^{-im\sigma}
  \;=\;
  \left((R_+)^{-1}\right)^\prime\of{\sigma}
  \mathcal{P}^\mu\of{(R_+)^{-1}\of{\sigma}}
  \,.
\end{eqnarray}

This holds true on $\mathbf{P}_k$ \refer{nice part of phase space} where we can use the fact
that $R_\pm$ are invertible.

Comparison with \refer{reparameterized cal P} shows that these are just 
reparameterizations
of the original local worldsheet fields $\mathcal{P}_\pm^\mu$, albeit with
a reparameterization that varies from phase space point to phase space point,
which is crucial for their invariance. But because the proof
\refer{proof of finite invariance} of \refer{finite invariance} 
involves only data available at a
single point in phase space,
it follows that for \emph{every} invariant expression
$F\of{\mathcal{P}_-}$ of the worldsheet fields $\mathcal{P}_-^\mu$ with
$\commutator{L_m}{F\of{\mathcal{P}_-}} = 0\,,\forall m$ we have
\begin{eqnarray}
  F({\mathcal{P}_-}) &=& F({\mathcal{P}^R_-})
\end{eqnarray}
(on $\mathbf{P}_k$), and analogously for $\mathcal{P}_+$.

\emph{In summary} we therefore obtain the following result:

On the restricted phase space $\mathbf{P}_\mathrm{L}$
\refer{nice part of phase space}
the classical Pohlmeyer invariants \refer{definition Pohlmeyer invariants} 
can be expressed in terms of
the classical DDF invariants 
\refer{definition classical DDF}
and the relation is 
\begin{eqnarray}
  \label{relation Pohlmeyer-DDF}
  Z^{\mu_1\cdots \mu_N}\of{\mathcal{P}}
  &=&
  Z^{\mu_1\cdots \mu_N}(\mathcal{P}^R)
  \,,
\end{eqnarray}
where $\mathcal{P}$ is the ordinary worldsheet field \refer{chiral CCR},
and $\mathcal{P}^R$ is the linear combination 
\refer{local pre-invariants} of classical DDF observables.

This can be expressed in words also as follows: The Pohlmeyer invariants
are left intact when replacing oscillators by respective DDF observables
in their oscillator expansion ($\alpha_m^\mu \to A_m^\mu$
, $\tilde \alpha_m^\mu \to \tilde A_m^\mu$).
Note that the Pohlmeyer invariants are all of level 0 in the sense of
\refer{level matching condition} so that the level matching condition is
trivially satisfied.

Because every polynomial in the DDF observables 
is consistently quantized by replacing $A_m^\mu$ and $\tilde A_m^\mu$ by the
respective operators discussed in \S\fullref{DDF operators}, this 
yields a consistent quantization of the Pohlmeyer invariants.\\

Finally, by simply generalizing the DDF invariants to the superstring,
equation \refer{relation Pohlmeyer-DDF} defines the generalization of the
Pohlmeyer invariants to the superstring. This is discussed in the 
next subsection:

\subsubsection{DDF and Pohlmeyer invariants for superstring}
\label{DDF and Pohlmeyer invariants for superstring}

The additional fermionc fields on the classical superstring shall here be 
denoted by $\Gamma_\pm^\mu\of{\sigma}$, which are taken to be normalized
so that their fermionic Poisson bracket reads
\begin{eqnarray}
  \antiCommutator{\Gamma_\pm^\mu\of{\sigma}}{\Gamma^\nu_\pm\of{\kappa}}_\mathrm{PB}
  &=&
  \pm 2\eta^{\mu\nu}\delta\of{\sigma-\kappa}
  \nonumber\\
  \antiCommutator{\Gamma_\pm^\mu\of{\sigma}}{\Gamma^\nu_\mp\of{\kappa}}_\mathrm{PB}
  &=& 0
  \,.
\end{eqnarray}
The modes are of course
\begin{eqnarray}
  b_r^\mu 
  &\defas&
  \frac{i}{\sqrt{4\pi}}
  \int e^{-ir\sigma}
  \Gamma_-^\mu\of{\sigma}
  \nonumber\\
  \tilde b_r^\mu
  &=&
  \frac{1}{\sqrt{4\pi}}
  \int e^{+ir\sigma}
  \Gamma_+^\mu\of{\sigma}  
\end{eqnarray}
with non-vanishing brackets
\begin{eqnarray}
  \antiCommutator{b^\mu_r}{b^\nu_s}_\mathrm{PB}
  &=&
  -i
  \eta^{\mu\nu}
  \delta_{r+s,0}
  \nonumber\\
  \antiCommutator{\tilde b^\mu_r}{\tilde b^\nu_s}_\mathrm{PB}
  &=&
  -i
  \eta^{\mu\nu}
  \delta_{r+s,0}  
  \,,
\end{eqnarray}
and the fermionic part of the super Virasoro constraints are
\begin{eqnarray}
  G_r &\defas&
  \frac{i}{\sqrt{2}}
  \int e^{-ir\sigma}
  \Gamma_-\of{\sigma} \inner \mathcal{P}_-\of{\sigma}
  \;d\sigma
  \;=\;
  \sum\limits_{m=-\infty}^\infty
  b_{r+m}\inner \alpha_{-m}
  \nonumber\\
  \tilde G_r 
  &\defas&
  \frac{1}{\sqrt{2}}
  \int e^{+ir\sigma}
  \Gamma_+\of{\sigma} \inner \mathcal{P}_+\of{\sigma}
  \;d\sigma
  \;=\;
  \sum\limits_{m=-\infty}^\infty
  \tilde b_{r+m}\inner \tilde \alpha_{-m}
  \,.  
\end{eqnarray}
The point is that we can entirely follow the constructions discussed in \S\fullref{DDF operators} to
get classical DDF invariants $A_m^\mu$ and $B_m^\mu$ which Poisson-commute
with the full set of super Virasoro constraints. For instance in the R sector 
the DDF observable $A_m^\mu$ is
\begin{eqnarray}
  A_m^\mu
  &\defas&
  \antiCommutator{G_0}{\frac{i}{\sqrt{4\pi}}\int \Gamma_-^\mu\of{\sigma}e^{-im R_-\of{\sigma}}}
  \nonumber\\
  &=&
  \frac{1}{\sqrt{2\pi}}
  \int d\sigma\;
  \left(
    \mathcal{P}_-^\mu\of{\sigma}
    +
    im \frac{2\pi\sqrt{2T}}{k\inner p}\Gamma_-^\mu\of{\sigma} k\inner \Gamma_-\of{\sigma}
  \right)
  e^{-imR_-\of{\sigma}}
  \,.
\end{eqnarray}
By making the replacement $\alpha_m^\mu \to A_m^\mu$ in the ordinary Pohlmeyer
invariant $Z^{\mu_1\cdots \mu_N}\of{\mathcal{P}_-}$ one obtains an object
whose purely bosonic terms exactly coincide with the ordinary bosonic
Pohlmeyer invariant and which furthermore has fermionic terms such that it
super-Poisson-commutes with all super Virasoro constraints. This object
is therefore obviously the superstring generalization of the ordinary Pohlmeyer invariant
of the bosonic string.

\section{Summary and Conclusion}
\label{summary and conclusion}

It has been shown that and how the Pohlmeyer invariants of the
closed bosonic classical string can be expressed in terms of the
classical analogs of the well known DDF operators. Heuristically, the construction
is based on the observation that the DDF invariants are
nothing but `dynamically' reparameterized worldsheet oscillators and
that the Pohlmeyer charges, being invariant under reparameterizations,
remain unaffected under an exchange of ordinary oscillators with the
respective DDF invariants.

This observation has some immediate consequences for the quantization
program associated with the study of Pohlmeyer-invariants: A quantization of the
DDF invariants in the usual way, leading to the DDF operators, is, by the
above result, also a quantization of the Pohlmeyer invariants and their algebra.
In particular this quantization is consistent in the sense that the commutator
of two quantized Pohlmeyer invariants is itself again an invariant, simply because
the algebra of DDF operators closes. This result should hence help to clarify
some questions of the Pohlmeyer program which have so far remained open
\cite{MeusburgerRehren:2002,Bahns:2004}. 

But above that the understanding of the relation between DDF and Pohlmeyer invariants
allows to immediately generalize the latter to the superstring and indeed to all
two-dimensional superconformal field theories for which DDF operators can be written
down. This in particular includes (super)strings on backgrounds which allow a free field
realization of the worldsheet fields, such as (super)strings on pp-wave backgrounds
\cite{Hikida:2003}.\\

From the point of view that the Pohlmeyer invariants are already included in the algebra of the
common and well-known DDF observables one may wonder if they deserve any special attention at all.

In this respect we noted that the relation of the Pohlmeyer invariants to 
Wilson loops of constant large-$N$ gauge connections along the string is intriguingly
reminiscent of the way states of string are expressed in terms of similar Wilson lines
in the IIB(IKKT) matrix model. Maybe this points to an interesting relationship yet to
be understood.

Another interesting aspect of the Pohlmeyer invariants is that they are all 
necessarily of vanishing oscillator level, quite contrary to the \emph{generic} 
DDF invariant, and that they still form a \emph{complete} set of invariant observables
\cite{Pohlmeyer:1988} in the sense that the knowlege of their classical values allows 
to locally reconstruct the string's worldsheet.

The vanishing of the level number of the Pohlmeyer invariants has the
maybe interesting consequence that for the superstring they are linear combinations
of terms of the form
$\commutator{G_{-\nu}}{c_1}\commutator{G_{-\nu}}{c_2}\cdots\commutator{G_{-\nu}}{c_p}$,
(where $\nu = 1/2$ in the NS and $\nu = 0$ in the R sector and similarly for the
other chirality sector), without any correction terms containing the coordinate
field 0-modes (\cf \refer{DDF invariants}.) Since $G_{-\nu}$ is a Dirac
operator (the Dirac-Ramond operator) on loop space, this is essentially an exact differential
form in the sense of Connes' noncommutative (spectral) geometry (e.g. \cite{Varilli:1997}).
Related observations have been made in section 4.4 of \cite{Schreiber:2004} and might point to 
an interesting meaning of the super-Pohlmeyer invariants, which has not fully emerged 
yet.\\

\paragraph{Note:}
After this work was completed we learned of the old articles
\cite{BorodulinIsaev:1982,Isaev:1983} where essentially the same results as given here are already reported.
Their relevance for the Pohlmeyer program and for attempts at ``alternative'' 
quantizations of the string seems not to have been widely familiar \cite{Schreiber:2004h}.

\acknowledgments{
  I am grateful to Robert Graham for his support and to
  K.-H. Rehren for detailed discussions. I would also like to
  acknowledge interesting discussions with H. Nicolai and D. Bahns.
  Finally many thanks to A. P. Isaev for making me aware of his work.

This work has been supported by the SFB/TR 12.
}
\newpage
\bibliography{std}

\end{document}